\documentclass[aps,twocolumn]{revtex4}
\usepackage{graphicx,amsmath,amsfonts,amssymb}

\newcommand{\ww}{\omega_\mathrm{d}}
\newcommand{\wo}{\omega_0}
\newcommand{\w}{\omega_\mathrm{b}}

\begin{document}

\title{{\bf Discrete breathers in a forced-damped array of coupled pendula:
Modeling, Computation and Experiment}}
\author{J. Cuevas$^1$, L.Q English$^2$, P.G. Kevrekidis$^3$, M. Anderson$^4$}
\affiliation{$^{1}$Grupo de F\'{\i}sica No Lineal, Departamento de \'{F}isica Aplicada I,
Escuela Universitaria Politecnica, C/ Virgen de \'{A}frica, 7, 41011
Sevilla, Spain \\
$^{2}$Department of Physics \& Astronomy, Dickinson College, Carlisle, PA 17013, USA \\
$^{3}$Department of Mathematics and Statistics, University of Massachusetts,
Amherst, MA 01003-4515, USA}
\date{\today}

\begin{abstract}
In this work, we present a mechanical example of an {\it experimental
realization} of a stability reversal between on-site and inter-site
centered localized modes. A corresponding realization of a
vanishing of the Peierls-Nabarro barrier allows for an experimentally
observed enhanced mobility of the localized modes near the reversal point.
These features are supported by detailed numerical computations of the
stability and mobility of the discrete breathers in this system of
forced and damped coupled pendula. Furthermore, additional exotic
features of the relevant model, such as dark breathers are briefly discussed.
\end{abstract}

\maketitle

{\it Introduction.} Discrete breathers (also referred to as intrinsic
localized modes or ILMs) have been the focal point of numerous
studies over the past two decades (for a recent review of the
relevant activity see e.g. \cite{FG08}). Such localized modes of
nonlinear lattices have been found to be fairly generic and constitute
a new paradigm in nonlinear science whose applications extend to
physics and chemistry, as well as to biology. Their original
proposal in the broadly applicable context of
anharmonic lattices \cite{st88,pa90} and subsequent theoretical analysis
(including the rigorous proof of their existence under appropriate
conditions \cite{macaub94})
has led to numerous experimental realizations ranging from
optical waveguides and photorefractive crystals to micromechanical
cantilever arrays and Josephson junctions, as well as in Bose-Einstein
condensates and layered antiferromagnets, among many others
\cite{cfk04,FG08}.

Typically, these ILMs come into two principal varieties, namely
the on-site centered modes \cite{st88} and the inter-site centered
ones \cite{pa90}. In the simplest ones among the relevant models
supporting such structures, either the former, or the latter modes
are stable, and this property is maintained when
relevant parameters, such as the lattice
spacing, are varied \cite{KK}. However, more recently it has been illustrated
that stability exchanges are theoretically possible between these
two types of modes \cite{hadz}, which result either in a true \cite{alanc}
or approximate \cite{chong} vanishing of the, so-called, Peierls-Nabarro
barrier, namely the energy difference between these modes. This vanishing
is, in turn, relevant in that it enables the localized modes to move
along the lattice without facing such an energy barrier, thereby
enhancing their mobility \cite{hadz,alanc}.

Our aim in the present work is to demonstrate an {\it experimental
realization} of a setting which enables such {\it stability reversals}
between on-site and inter-site modes, namely  an array of pendula under
the action of driving and damping. A detailed analysis illustrates
such stability exchanges to occur numerically, and the experimental
findings validate this result. Furthermore, enhanced mobility of
the localized modes is observed experimentally and computationally
near the exchange point and additional theoretically interesting features
of the model are predicted (in a regime that is not currently accessible
experimentally). The structure of our presentation is as follows. In
the next section, we explain the experimental setup, followed by the
theoretical/numerical setup. We close the paper with a detailed discussion
of the results, as well as some potential future directions.

\begin{figure}
\begin{center}
\includegraphics[width=8cm]{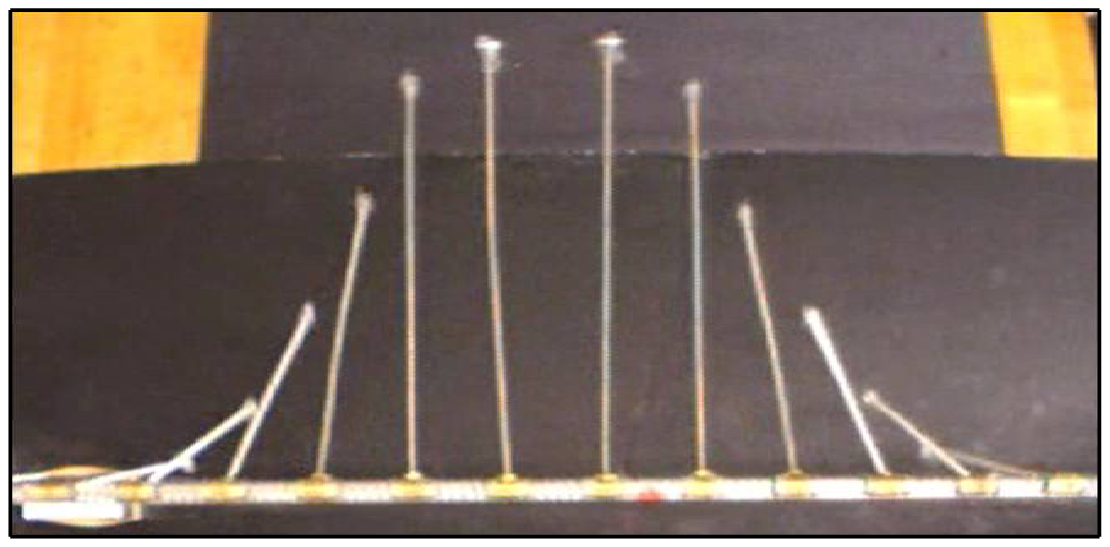}
\begin{tabular}{cc}
    \includegraphics[width=4cm]{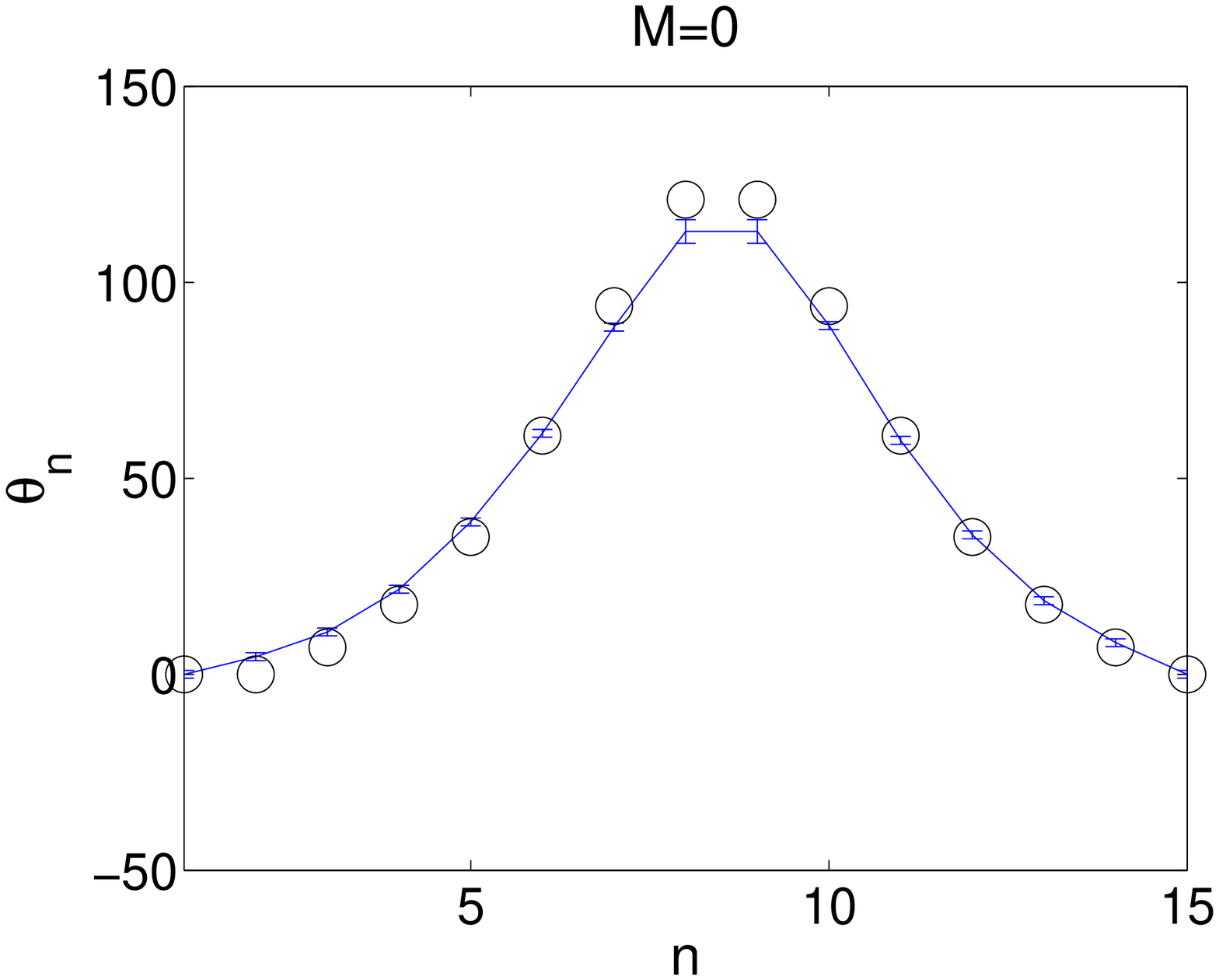} &
    \includegraphics[width=4cm]{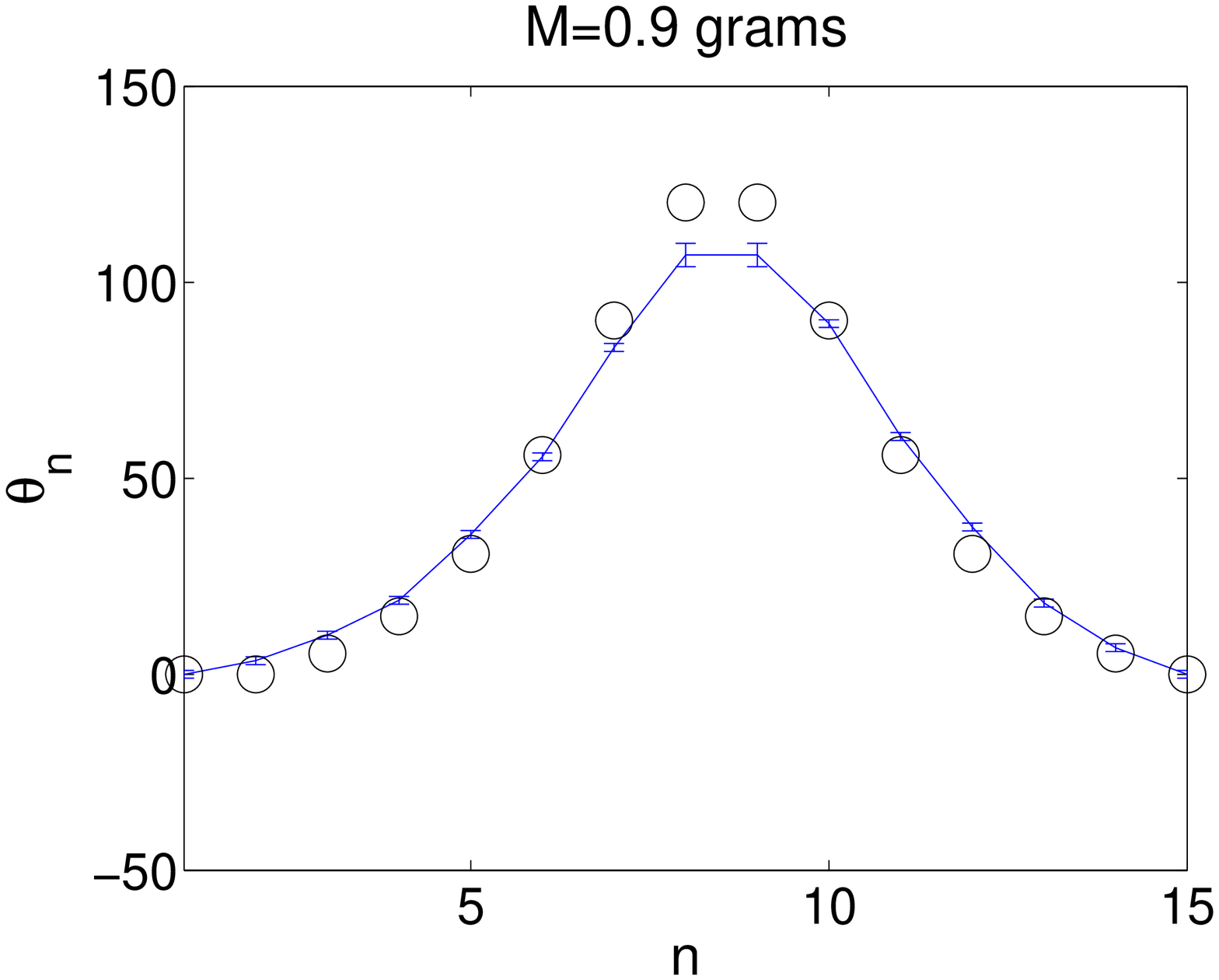} \\
    \includegraphics[width=4cm]{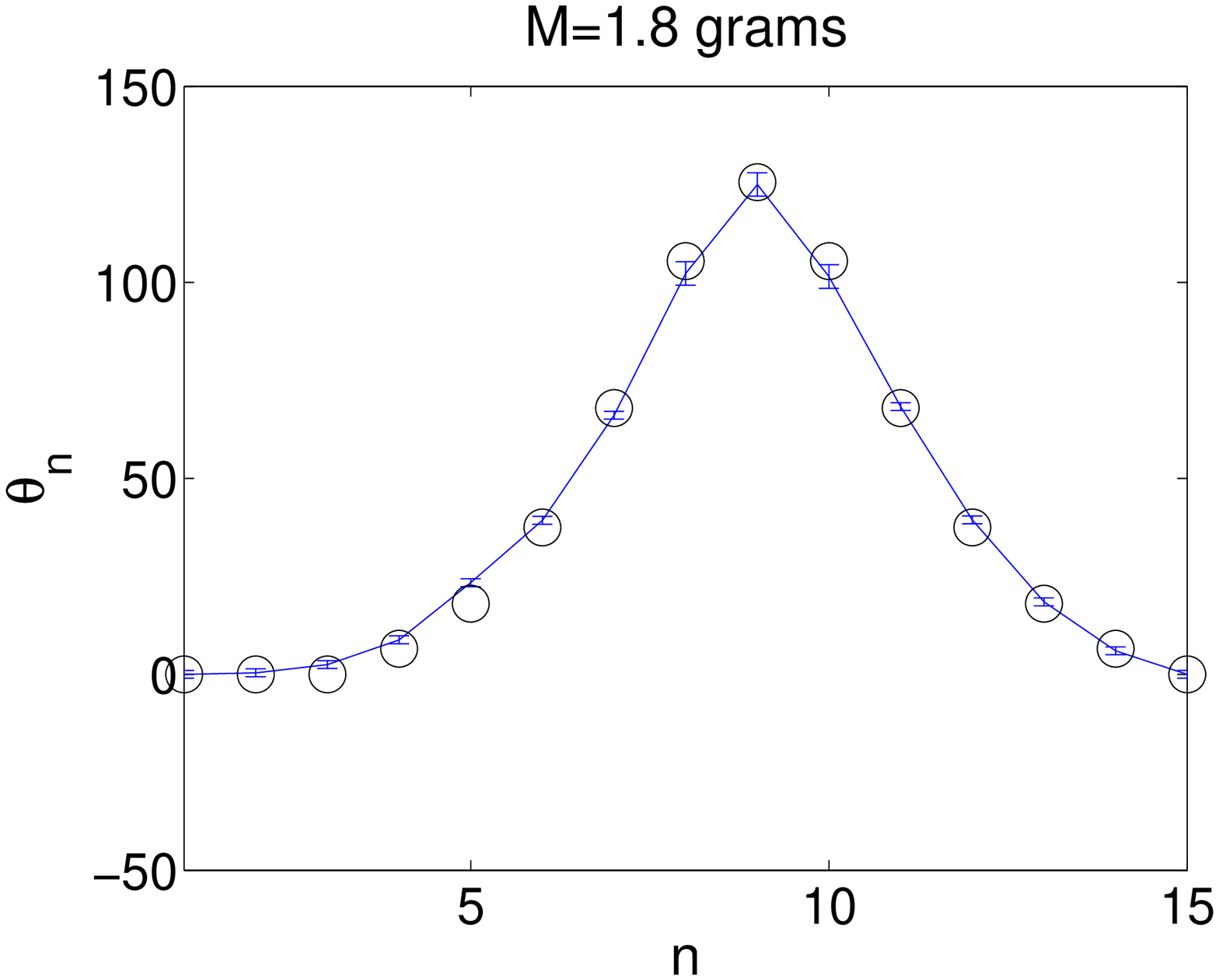} &
    \includegraphics[width=4cm]{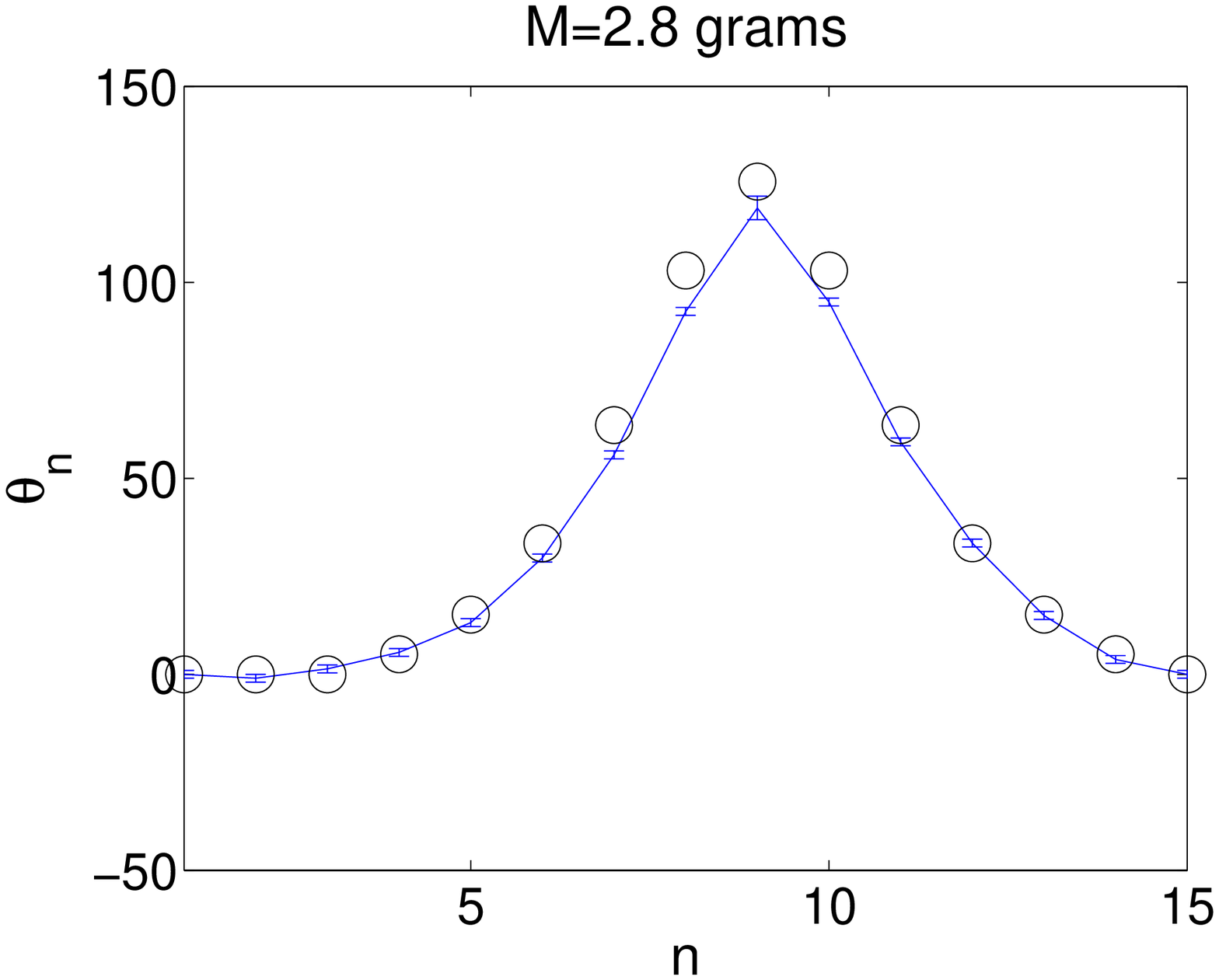} \\
\end{tabular}
\caption{(Color Online). Snapshot illustrating the experimental setup and
the observation of an inter-site breather (top panel), and comparison of experimental and numerical profiles of stable inter-site breathers (middle panels) and on-site breathers (bottom panels). In all cases, $A=1.12$ cm and $\w=0.87$. Circles represent the numerical results whereas the full lines with error bars correspond to the experimental profiles.} \label{fig:profiles}
\end{center}
\end{figure}

\begin{figure}
\begin{center}
\begin{tabular}{cc}
    (a) & (b) \\
    \includegraphics[width=4cm]{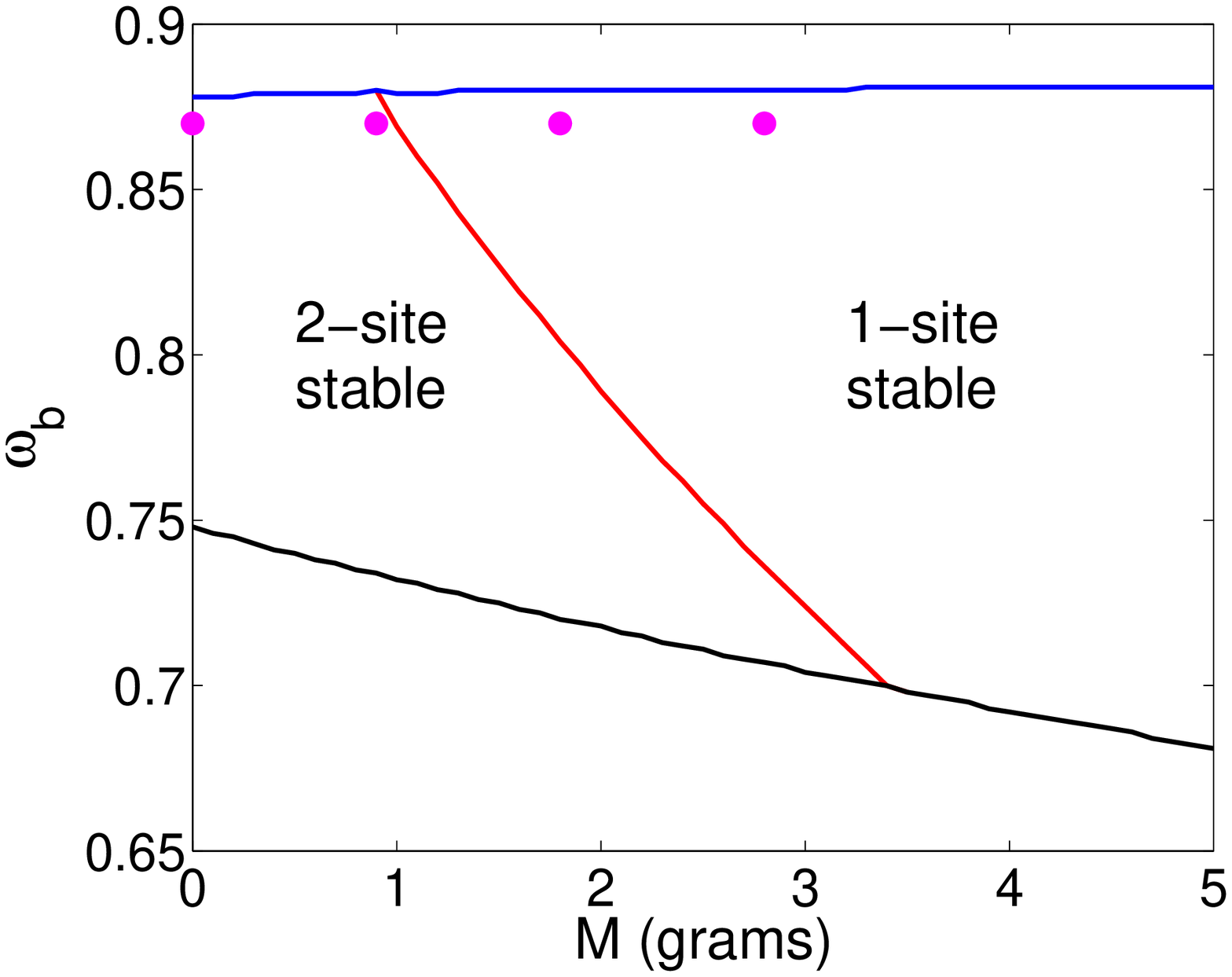} &
    \includegraphics[width=4cm]{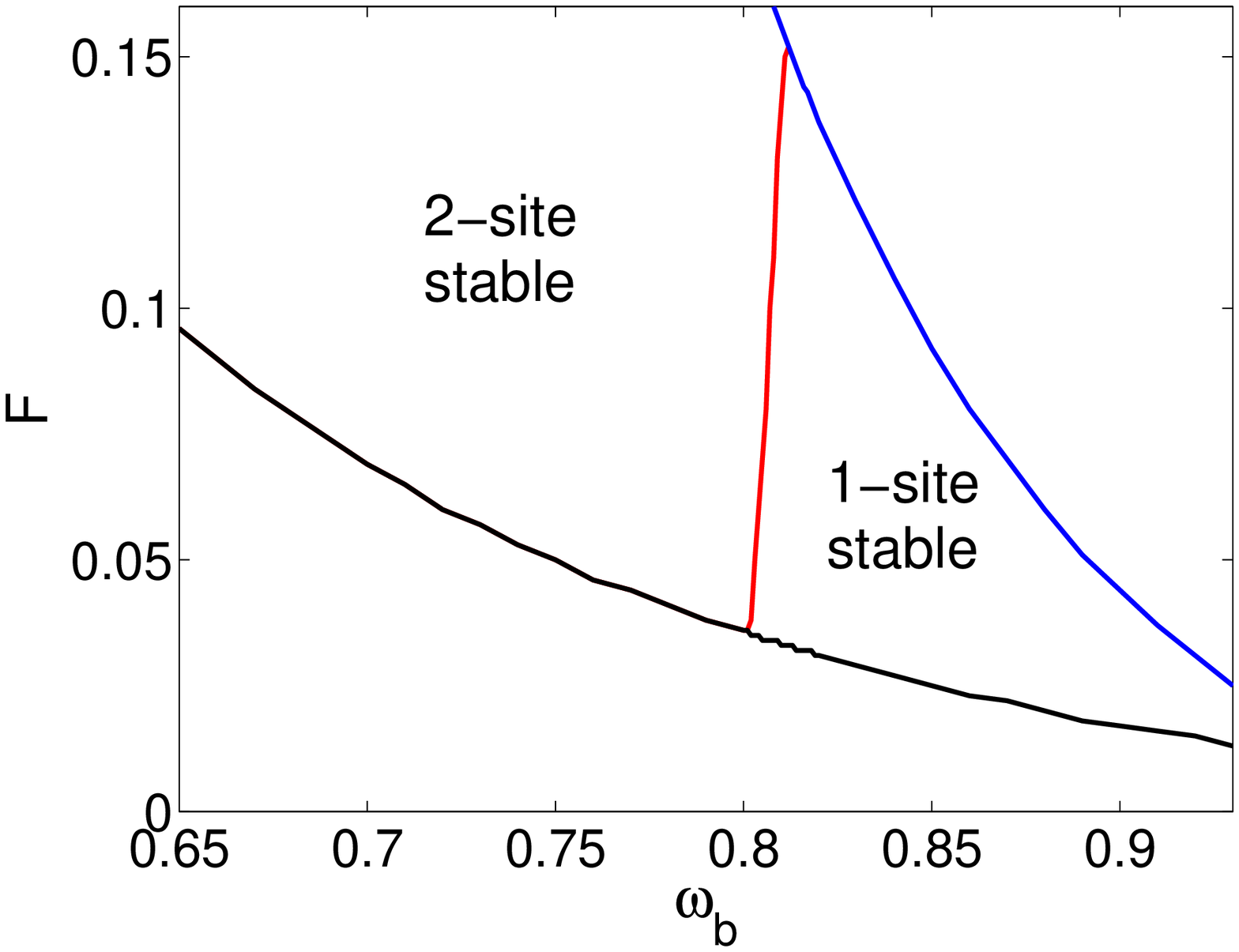} \\
    (c) & (d) \\
    \includegraphics[width=4cm]{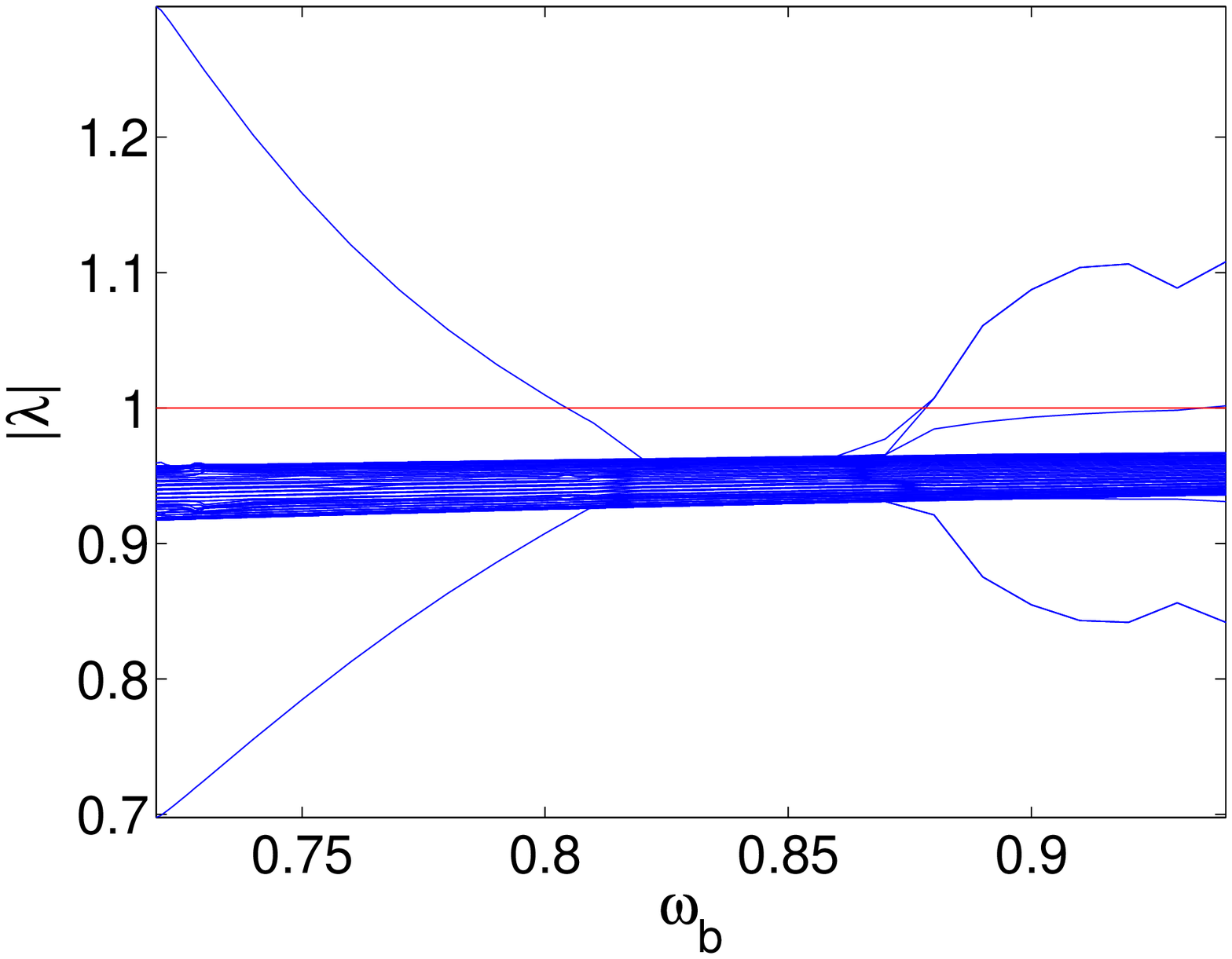} &
    \includegraphics[width=4cm]{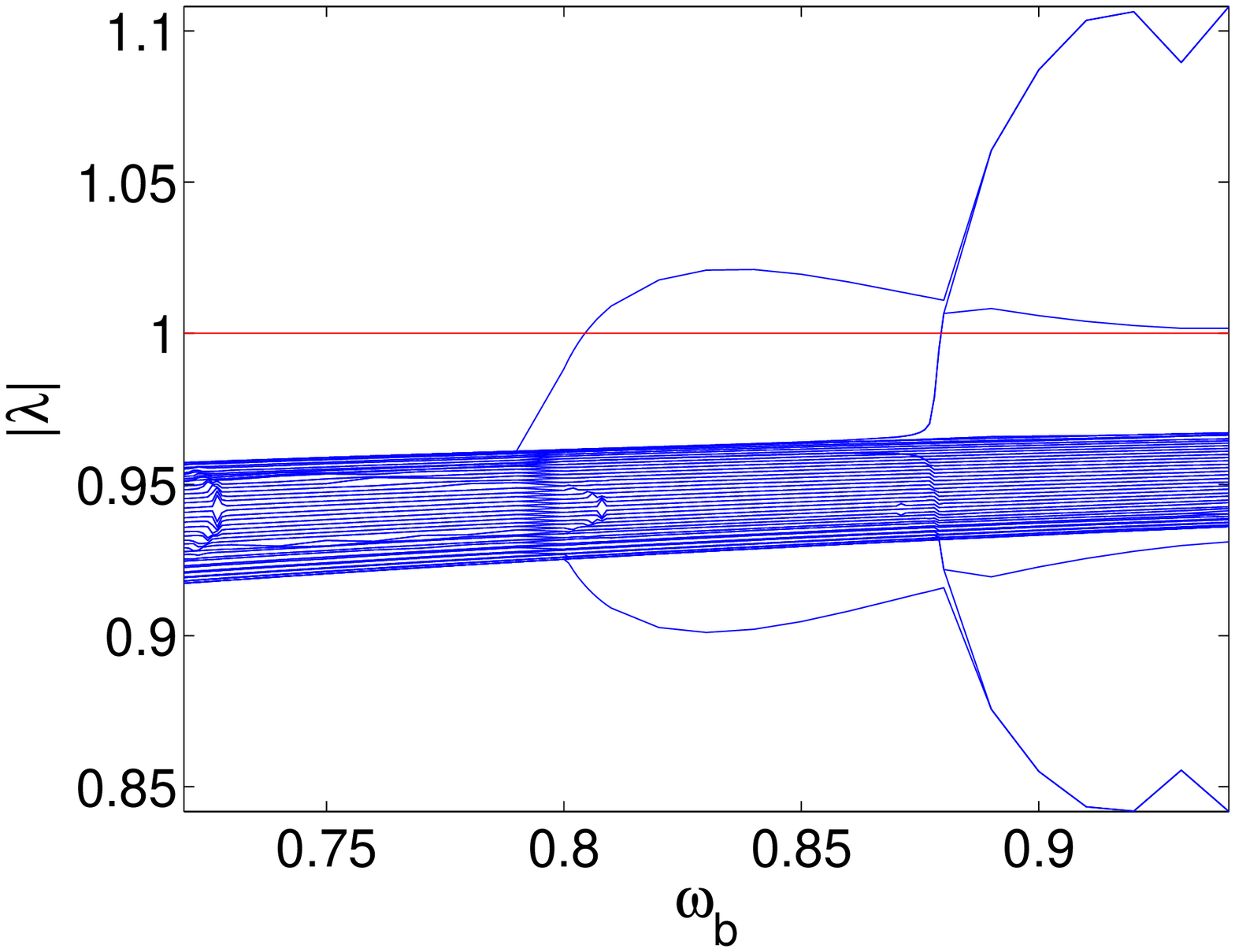} \\
\end{tabular}
\caption{(Color Online). Top: 2-parameter planes for (left) fixed $A=1.12$ cm and (right) fixed $M=1.8$ grams. Full circles in the left panel correspond to the solutions shown in Fig. \ref{fig:profiles}. Bottom: moduli of Floquet multipliers (indicating stability, when they are all smaller than $1$) for (left) 1-site and (right) 2-site breathers. In this case, $F=0.06$ and $M=1.8$ grams are fixed, while $\w$ is variable. An exchange of stability bifurcation is observed
for $\w\in(0.804,0.805)$. In both cases, the bifurcation arising for higher $\w$ is a Naimark--Sacker one.}
\label{fig:planes}
\end{center}
\end{figure}


{\it Experimental Setup.} We investigate a driven and damped chain of pendula
each comprised of a rod of mass $m$ and length $L$ ending in a weight of mass $M$. They are suspended along a line by a taut piano wire stretched across a metal frame and connected via springs, as in  \cite{Sievers}. The system is driven sinusoidally along a horizontal direction with an adjustable amplitude $A$ and a tunable frequency $\omega_d$. The resulting dynamics of the pendulum chain is recorded by an overhead camera, and individual frames can be used to extract pendulum angles.
The governing equations of motion for a pendulum in the chain take into account the detailed effects of driving and damping:

\begin{eqnarray}\label{eq:dyn0}
\ddot\theta_n&+&\omega_0^2\sin\theta_n
-\left(\frac{\beta}{I}\right) \Delta_2 \theta_n
+ \frac{\gamma_1}{I}\dot\theta_n \nonumber \\
&-& \frac{\gamma_2}{I} \Delta_2 \dot{\theta}_n
+F\ww^2\cos(\ww t)\cos(\theta_n) = 0
\end{eqnarray}

\noindent with $I=ML^2+\frac{1}{3}mL^2$, $\wo^2=\frac{1}{I}\left(mg\frac{L}{2}+MgL\right)$ and    $F=\frac{A\wo^2}{g}$, while $\Delta_2$ represents the discrete Laplacian. Here, $g$ is 9.80 N/kg, and $\beta$ denotes the torsional spring constant. The nature of damping in the system warrants closer inspection. Two types of friction are present in the experiment - an on-site and an inter-site
one. The on-site component (of prefactor $\gamma_1$) comes from air drag as well as contact friction at the suspension points. The inter-site component
(of prefactor $\gamma_2$)
has its origin in the energy dissipation due to the twisting of the springs,
and it should be related to angle differences between neighboring springs.
This draws interesting parallels to the lattice turbulence models of \cite{peyrard} and recent models
of granular crystals \cite{granular}.
Equation (\ref{eq:dyn0}) can be non-dimensionalized by introducing the following parameters: $\w=\frac{\ww}{\wo}$, $C=\frac{\beta}{I\wo^2}$, and $\alpha=\frac{\gamma}{I \wo}$.

This model is similar to
that of \cite{Marin}, but
here the driving term depends on the angular displacement at each lattice site
and on the driving frequency. The definition of $C$ suggests two experimental ways of adjusting this parameter. For gross adjustment, one can change the springs between the pendula modifying the torsional stiffness and coupling $\beta$. A finer (and more convenient) way of varying $C$ is via $I$ and $\wo$. This can be done most easily by changing the end-mass $M$ which affects $I$ and $\wo$
(see the respective definitions). Note that by changing $\wo$, $\w$ and $F$ are also affected.

The on-site damping constant, $\gamma_1$, was measured
using just one isolated pendulum, tracking its amplitude decay over time (which yields $\gamma_1/I$). Incidentally, $\gamma_1$ was measured for a number of endmasses, $M$, and found to be approximately independent of endmass for $M<10$g.

The parameter $\gamma_2$ was estimated as follows. Starting with a group of three pendula connected in a line via springs, the two outer ones are held fixed vertically, and the amplitude decay of the center pendulum is tracked. The exponential decay constant thus obtained is then divided by the angular velocity $\omega$ for this mode (in the linear regime), which can be measured independently. This number corresponds to an effective $\alpha$, and it is larger than $\alpha_1=\gamma_1/(I \omega_0)$.

Table \ref{values} gives the measured values of the experimental parameters.
The driver amplitude, $A (=1.12{\rm cm})$, was selected to yield a value $F$ of around 0.06.

\begin{table}[h]
\caption{Experimental constants}
\label{values}
\begin{center}
\begin{tabular}{|c|c|c|c|c|c|}
\hline
$\beta $ (Nm/rad) & $\gamma_1$ (g cm$^2$/s) & $\gamma_2$ (g cm$^2$/s) & $L$ (cm) & $m$ (g) & $A$ (cm)\\
\hline \hline
0.0165 & 284 & 70  & 25.4 & 13  & 1.12  \\
\hline
\end{tabular}
\end{center}
\end{table}

{\it Theoretical/Numerical Setup.}
We consider the breather existence and stability in the (non-dimensional)
equations of motion.
Our computations start at the anti-continuous (AC) limit ($C=0$), using the
two (different amplitude) attractors of an uncoupled pendulum. Breathers close to the AC limit only exist if there are two coexisting attractors of the single driven and damped pendulum - both phase-locked to the external driving.
The highest-amplitude attractor will correspond to the excited
breather site(s) whereas the smallest-amplitude one will be chosen for the
low-amplitude background. Then, the breather will be a collective oscillation
at the driving frequency, $\w$. For the analyzed parameter
range, the attractors oscillate in anti-phase. As
a consequence, the excited site of the breather oscillates in anti-phase with
respect to the tails, in agreement with the experimental observations
of \cite{Sievers}.
Upon creation at the AC limit, such solutions have been continued for
arbitrary $C\neq0$, through a path-following scheme in the real space,
as proposed in \cite{MA96}; this method allows us to indentify both stable
and unstable solutions.

The linear stability of the ensuing discrete breathers is studied by means of
a Floquet analysis. To this end, we add a small perturbation $\xi_n$ to the
breather solution leading to the following equation for the perturbations:

\begin{equation}\label{eq:stab1}
    \ddot\xi_n+\alpha\dot\xi_n+g(t;\theta_n(t))\xi_n -C \Delta_2 \xi_n
    -\alpha_2 \Delta_2 \dot{\xi}_n=0
\end{equation}

\noindent with $g(t;\theta_n(t))=\cos \theta_n(t)-F\w^2\cos(\w t)\sin\theta_n(t)$

Stability analysis is performed by diagonalising the monodromy matrix ${\mathcal M}$, which relates the perturbation at $t=0$ to that at $t=T$:
\begin{equation}
    \begin{pmatrix}
      \xi_n(T) \\
      \dot \xi_n(T) \\
    \end{pmatrix}
    = {\mathcal M}
    \begin{pmatrix}
      \xi_n(0) \\
      \dot \xi_n(0) \\
    \end{pmatrix}
\end{equation}
The linear stability of breathers requires that the monodromy eigenvalues
(Floquet multipliers) must be inside (or at) the unit circle; see e.g.
\cite{Aubry,MST} for details. Due to the inter-site damping, only one
property (implied by the real character of the monodromy) can be extracted
for the multipliers, which is that they appear in complex conjugated pairs
($\lambda$,$\lambda^*$).

\begin{figure}
\begin{center}
\begin{tabular}{cc}
    \includegraphics[width=4cm]{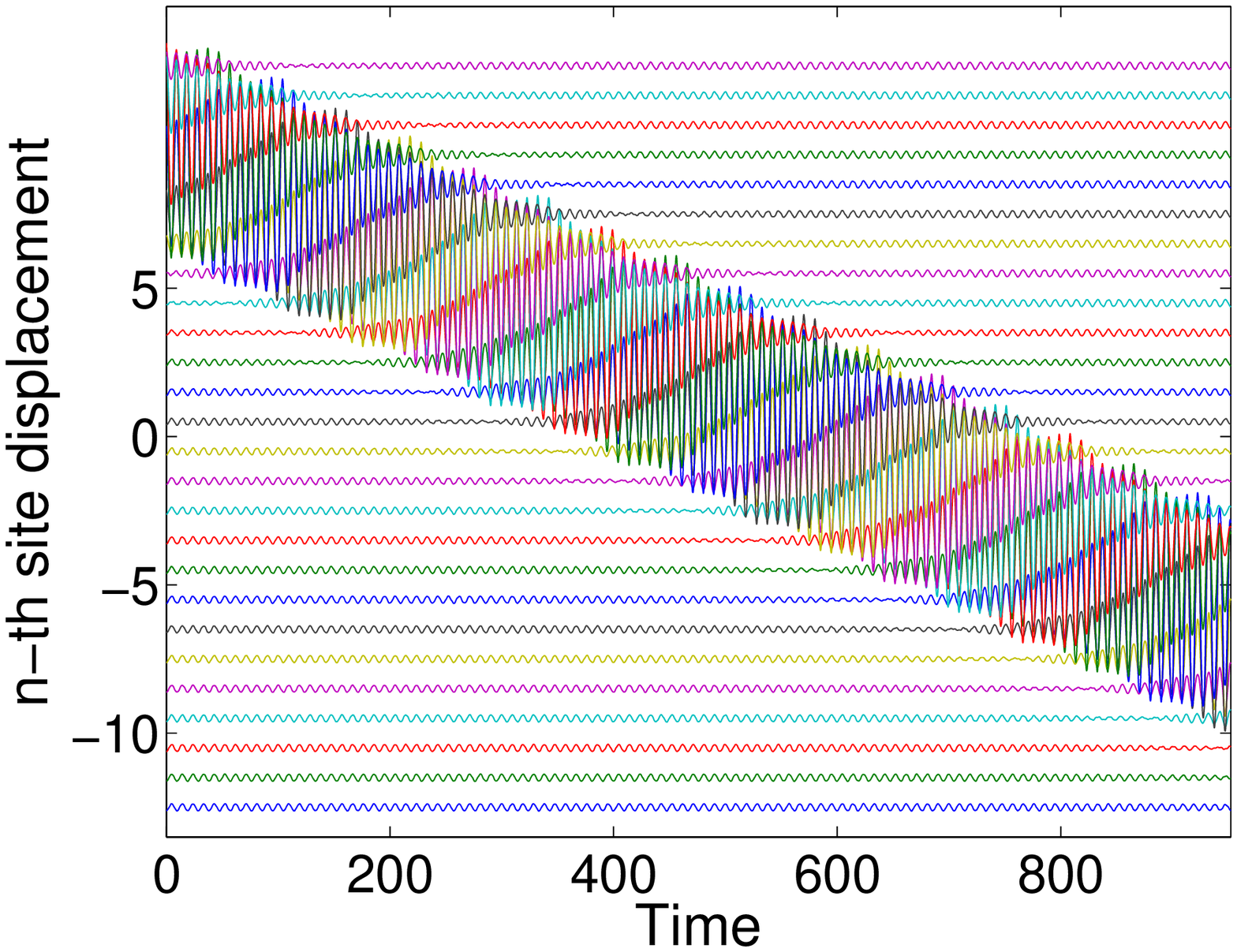} &
    \includegraphics[width=4cm] {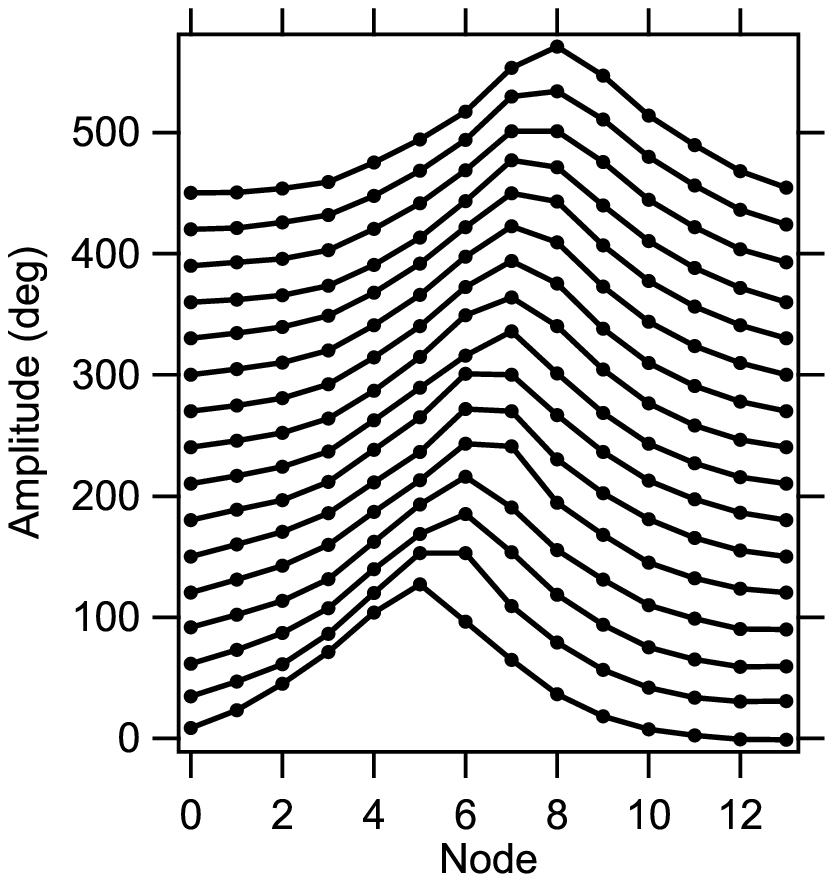} \\
\end{tabular}
\caption{(Color Online). (Left panel) Numerical  (for $F=0.15$) and (right panel)
experimental examples of breather mobility. In the left panel, $M=3.8$ grams
and $\w=0.66$, while the right panel example is for $M=1.8g$ and
$\omega_b=0.87$. The vertically stacked experimental
profiles are spaced one period apart.} \label{fig:moving}
\end{center}
\end{figure}

{\it Results and Discussion.}
Let us start by comparing typical experimental and numerical
breather profiles. Subsequently, we will explore the regions in parameter space where the two types of breathers are stable, again comparing theory and experiment. Finally, we briefly discuss
additional interesting system features such as moving and dark breathers.

In order to compare with experimental profiles, we choose a chain with 13
pendula and fixed-end boundary conditions. Fig. \ref{fig:profiles} details
the experiment-theory comparison. The circles indicate computed values and the line shows the experimental profile; experimental uncertainties are indicated by the error bars. The uncertainty for angles higher than 100 degrees is $\pm3$º whereas it is $\pm1$º if the angles are smaller. An exchange of stability is
identified {\it both} numerically and experimentally: for small masses,
2-site breathers are stable whereas for larger masses, 1-site breathers
become stable.

The experimentally accessible parameters are $M$, $\w$, and $F$ (via $A$). To explore
breather stability and symmetry more fully, we generate two-parameter
diagrams for a larger-sized array. To this end, we have chosen a chain of
$N=41$ pendula (so that boundary effects are of lesser importance).
Fig. \ref{fig:planes}(a) illustrates the breather stability for fixed
$A=1.12$ cm (left panel where $M$ and $\w$ are varied). We observe that
as $M$ is increased, at a given $\w$, an exchange of stability bifurcation is observed: the 1-site breather gains stability, and (practically) simultaneously the 2-site breather solution loses stability. Conversely, for a given $M$,
an exchange of
stability can be induced by varying $\w$. Above a critical value
($\approx 0.875$), the system experiences a Naimark-Sacker (NS) bifurcation
(viz. generalized Hopf or Hopf bifurcation of periodic orbits) leading to the
eventual breather disappearance.
For driver frequencies $\w$ below a critical lower value, we observe the
non-existence of any kind of breather (even unstable ones).
Notice that the four dots in Fig. \ref{fig:planes}(a) correspond to the
parameters of Fig. \ref{fig:profiles}) with M=0g and M=0.9g falling into the 2-site regime at $\w$=0.87, whereas M=1.8g and M=2.8g are in the 1-site regime.

According to the figure, the stability exchange can also be initiated
by varying the driver frequency at a fixed endmass, and this was also tested
experimentally. For M=1.8g, the 1-site ILM is found to be stable down to $\w\cong 0.83$. At that point, a transition to 2-site ILM appears to be initiated. The resulting 2-site ILM is, however, difficult to stabilize in the system of 13 pendula, and it usually decays within 5-10 periods. The observed transition frequency agrees reasonably well with the predicted value of 0.805. At M=1.2g, the transition can be observed more clearly. The on-site breather prevails at $\w$ = 0.96. Below $\w$ = 0.91 the inter-site breather is stable; the predicted value is, in this case, 0.852. The NS-bifurcation at larger $\w$ is also clearly observed experimentally but for endmasses
above M=6.8g (see also \cite{Sievers}).

In Fig. \ref{fig:planes}(b), a $F$-$\w$ plane for fixed $M=1.8$ g is depicted. Here we see that 2-site breathers are stable for small $\w$ with an exchange of stability bifurcation taking place in a narrow range of frequencies around $\w=0.81$. Below $\w=0.65$ breathers are NS-unstable.
Notice, again, that for small values of $F$ breathers can not be found. Figs. \ref{fig:planes}(c) and (d) show a typical parametric dependence of the Floquet exponents for two
different configurations.

The above stability exchanges between the two principal configurations
(and the corresponding vanishing of the energy barrier between them)
renders credible the scenario of breather mobility near such parameter
sets \cite{hadz,alanc}. Such mobility has been observed (even in fairly
large lattices with $\approx 100$ sites) and typically requires
appropriately large values of $F$. A typical example of such a traveling breather for $F=0.15$ is depicted in Fig. \ref{fig:moving} ($\w=0.66$ and $M=3.8$ g)
Such observations are in agreement with experimental indications of
mobility (albeit admittedly in the fairly short experimental lattice),
as shown in the bottom panel of the Fig. \ref{fig:moving}. The experimental parameters are M=1.8g (C = 0.8) and $\omega_b =  0.87$. In this regime, a stationary ILM can be made to move easily by bringing a fixed boundary into the wings of the ILM; in response, the ILM will move away from the boundary. The figure depicts an ILM repelled from the left boundary; it moves fairly quickly at first - alternating between 1-site and 2-site symmetry - then slows down and finally comes to rest three sites over.%
\begin{figure}[t]
\begin{center}
\begin{tabular}{cc}
    \includegraphics[width=4cm]{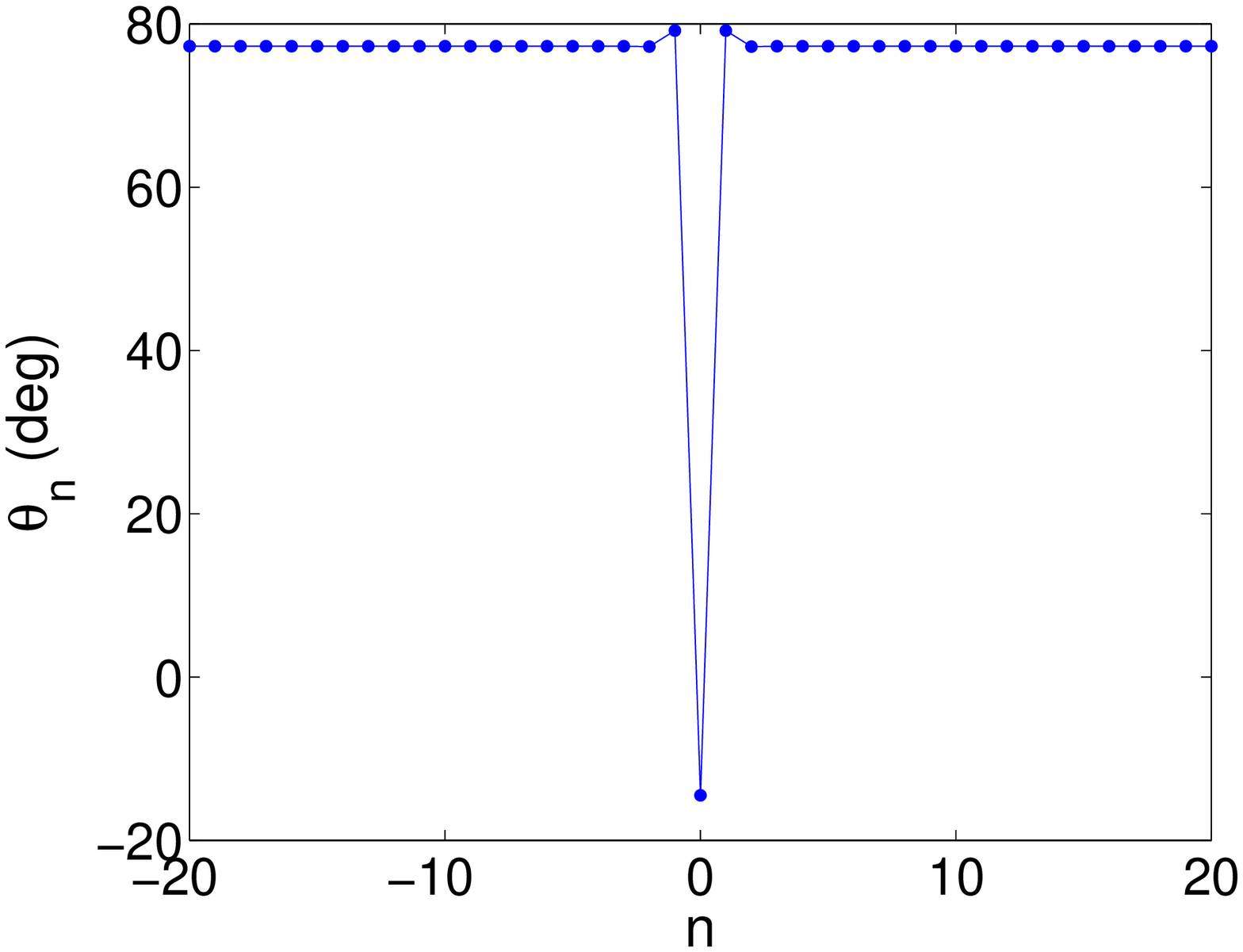} &
    \includegraphics[width=4cm]{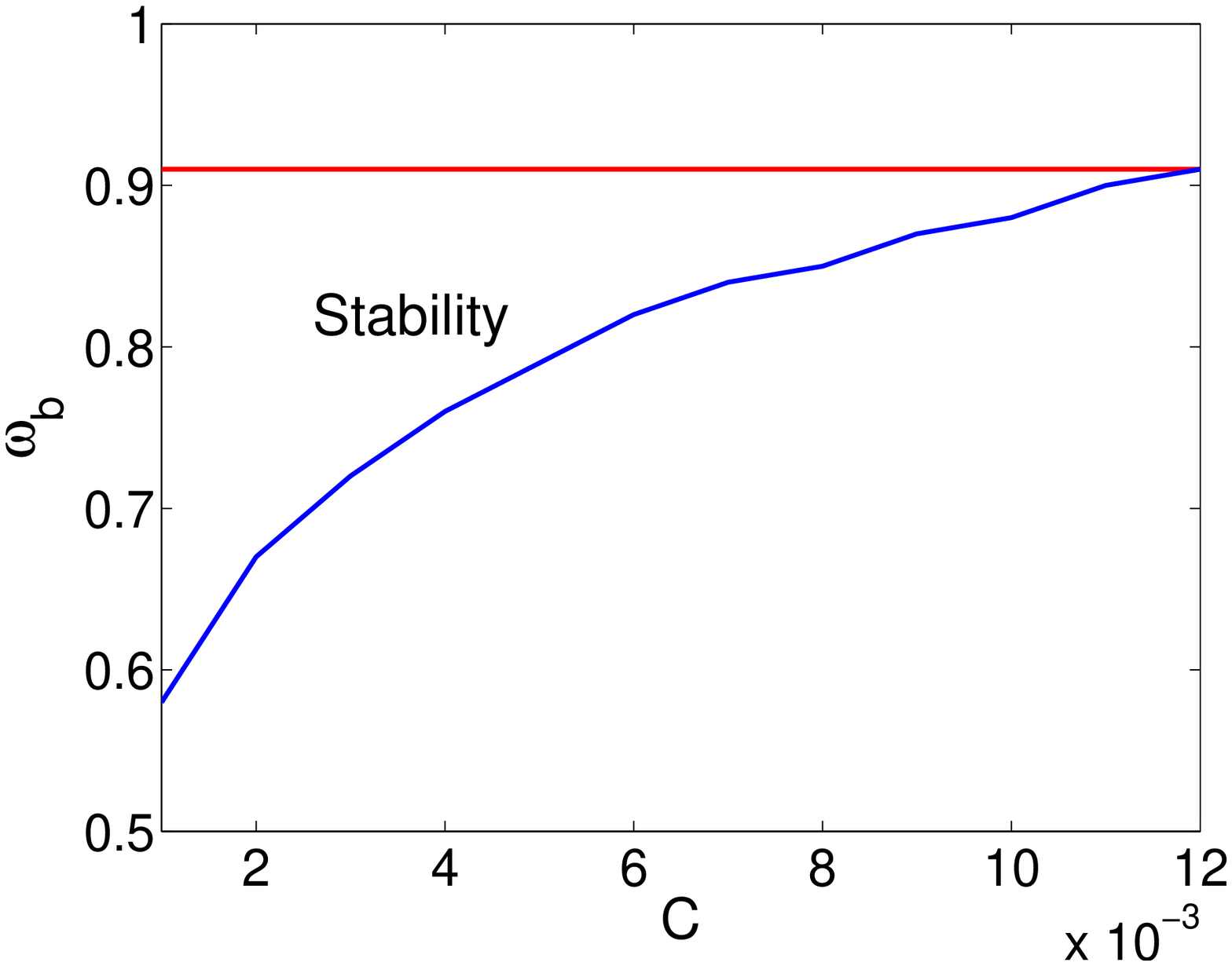} \\
    \includegraphics[width=4cm]{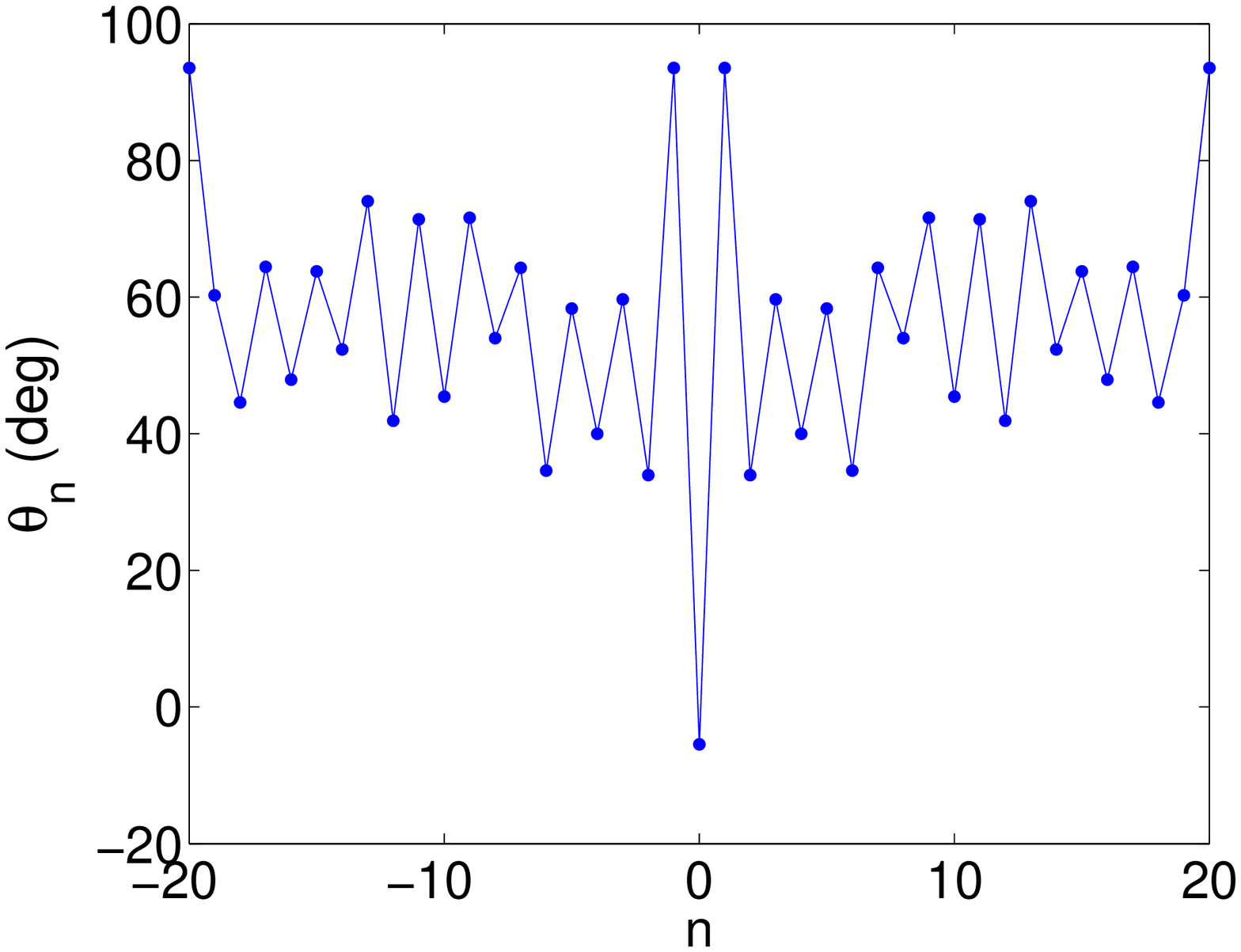} &
    \includegraphics[width=4cm]{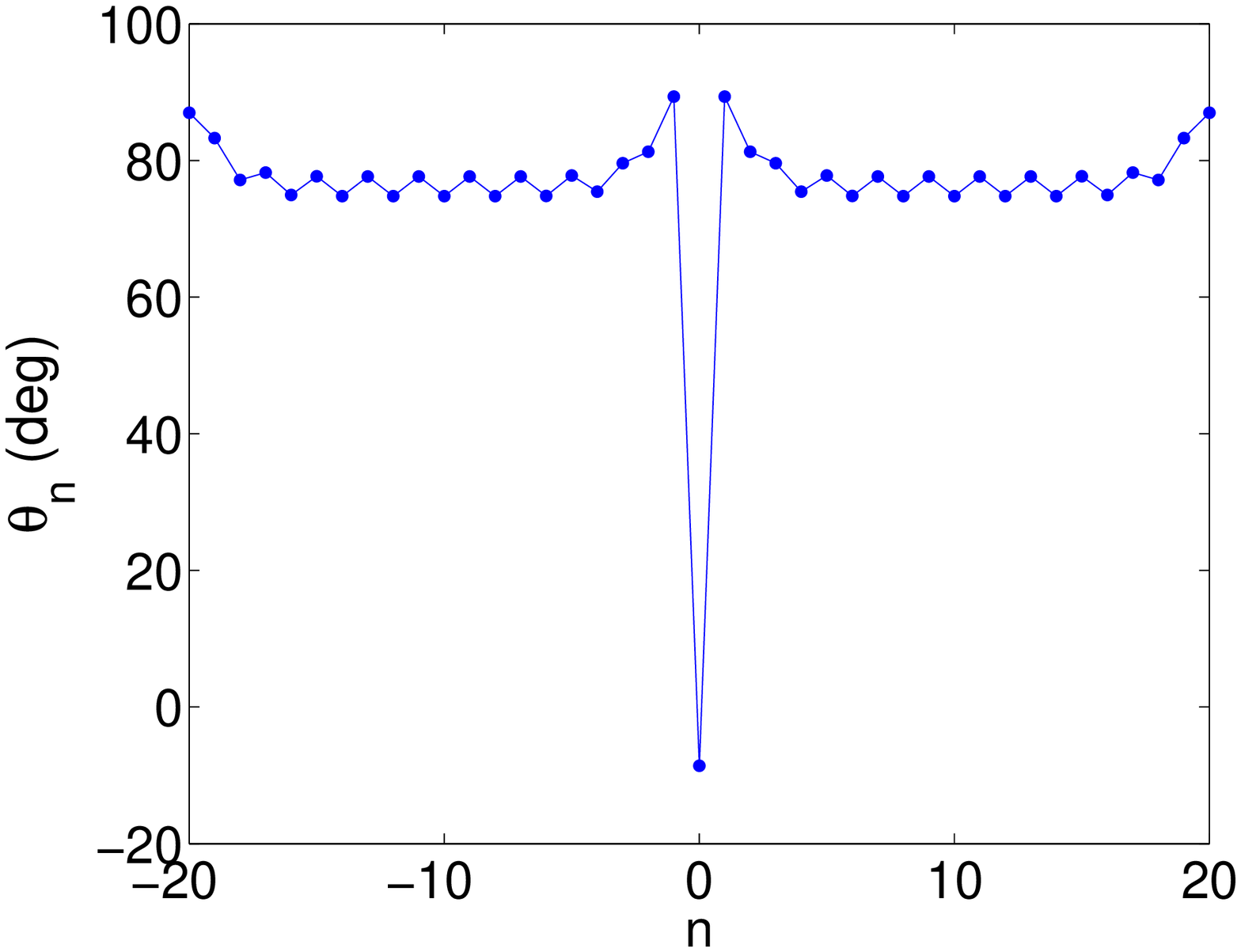} \\
\end{tabular}
\caption{(Color Online). (Top panels) Dark breather profile for $C=0.01$ and $\w=0.9$ (left) and $\w$-$C$ plane for the values specified in the text. (Bottom panels) Example of pseudo-dark breathers with $C=0.01$ which are stable in the range where ``real'' dark breathers are exponentially-unstable [(left) $\w=0.8$ and (right) $\w=0.86$].} \label{fig:dark}
\end{center}
\end{figure}
To demonstrate another interesting feature of the model, let us now
investigate the existence and stability of {\it dark breathers}.
In Ref. \cite{Dark} the existence of dark breathers in the dissipative Klein--Gordon lattice with non-parametric driving was demonstrated. Figure \ref{fig:dark} shows the existence and stability range of dark breathers. As shown, the range of stability of dark breathers is somewhat limited. We have varied parameters $\w$ and $C$ for fixed $\gamma=0.001$, $\gamma_2=0$ and $F=0.10$. For small $\w$, the dark breathers are exponentially unstable and lead to dark-localized structures.
For $\w$ above the range shown in the figure, the breathers experience a NS bifurcation. This instability is, however, small and system-size dependent.
It was numerically found that dark breathers do not exist for $\w\gtrsim0.95$, while
for $C>0.012$, stable dark breathers also do not exist - a constraint which
in our present experimental setting inhibits the formation of such coherent structures.

{\it Conclusions.} We have illustrated that the setting of damped
and driven coupled pendula is an ideal playground for showcasing
many interesting features of discrete breathers. In particular,
in addition to bearing unusual model features, such as inter-site
dissipation, this system exhibits complex stability properties evidenced by
stability exchanges between on-site and inter-site configurations. These, in turn,
lead to interesting breather dynamics, including the
potential for mobility of such localized excitations. Our numerical
results based on the proposed model were found to be in excellent qualitative
and good quantitative agreement with the
experimental findings. Determining whether multi-site excitations
or more exotic features such as dark breathers are possible would
be among the natural extensions of the present work.


\begin{thebibliography}{99}

\bibitem{FG08} S. Flach and A. Gorbach, Phys. Rep. {\bf 267}, 1 (2008).

\bibitem{st88} A.J. Sievers and S. Takeno,
Phys. Rev. Lett. {\bf 61}, 970 (1988).

\bibitem{pa90} J.B. Page,
Phys. Rev. B {\bf 41}, 7835 (1990).

\bibitem{macaub94} R.S. MacKay and S. Aubry, Nonlinearity {\bf 7},
1623 (1994).

\bibitem{cfk04} D.K. Campbell, S. Flach and Yu.S. Kivshar,
Phys. Today {\bf 57}, 43 (2004).

\bibitem{KK} T. Kapitula and P. Kevrekidis, Nonlinearity {\bf 14}, 533 (2001).

\bibitem{hadz} L. Hadzievski, A. Maluckov, M. Stepi{\'c} and D. Kip,
Phys. Rev. Lett. {\bf 93}, 033901 (2004).

\bibitem{alanc}  T.R. Melvin, A.R. Champneys, P.G. Kevrekidis
and J. Cuevas, Phys. Rev. Lett. {\bf 87}, 124101 (2006).

\bibitem{chong} R. Carretero-Gonz{\'a}lez, J.D. Talley, C. Chong
and B.A. Malomed, Phys. D {\bf 216}, 77 (2006).

\bibitem{Sievers} R Basu Thakhur, LQ English and AJ Sievers. J.
Phys. D: Appl. Phys. {\bf 41}, 015503 (2008).

\bibitem{peyrard} M. Peyrard and I. Daumont, Europhys. Lett.
{\bf 59}, 834 (2002).

\bibitem{granular} R. Carretero-Gonz{\'a}lez, D. Khatri, M.A. Porter,
P.G. Kevrekidis and C. Daraio, Phys. Rev. Lett. {\bf 102}, 024102 (2009).

\bibitem{Marin} JL Mar\'{\i}n, F Falo, PJ Mart\'{\i}nez and LM
Flor\'{\i}a. Phys. Rev. E {\bf 63}, 066603 (2001). PJ Mart\'{\i}nez, M Meister, LM Flor\'{\i}a and F Falo. Chaos {\bf 13}, 610 (2003) .

\bibitem{MA96} J.L.Mar\'{\i}n and S.Aubry. Nonlinearity {\bf 9}, 1501 (1996).

\bibitem{Aubry} S Aubry. Physica D {\bf 103}, 201 (1996).

\bibitem{MST} JFR Archilla, J Cuevas, B S\'{a}nchez--Rey and A
\'{A}lvarez. Physica D {\bf 180}, 235 (2003). J Cuevas, JFR Archilla
and FR Romero. Nonlinearity {\bf 18}, 769 (2005).

\bibitem{Dark} A \'{A}lvarez, JFR Archilla, J Cuevas and FR Romero.
New J. Phys. {\bf 4}, 72 (2002).

\end{thebibliography}
\end{document}